\documentclass{PoS}
\usepackage{graphicx}
\newcommand{\cd}{\makebox[0.08cm]{$\cdot$}}

\title{Field theory on the light-front}

\ShortTitle{Field theory on the light-front}

\author{\speaker{Jean-Fran\c cois Mathiot}%
\\
       Clermont Universit\'e, Laboratoire de Physique
Corpusculaire, BP10448, \\ F-63000 Clermont-Ferrand, France\\
       E-mail: \email{mathiot@clermont.in2p3.fr}}

\abstract{We present a general framework to study relativistic compound systems in a Hamiltonian formalism. This formalism is based on the explicitly covariant formulation of light-front dynamics, with a decomposition of the state vector in Fock components. In order to be able to make definite predictions order by order in the truncation of the Fock expansion, we use an appropriate Fock sector dependent renormalization scheme. Our covariant formulation enables us to have a strict control of any violation of rotational invariance due to the choice of a given orientation of the light-front plane. This is mandatory in order to define the necessary renormalization conditions, and hence to be able to calculate physical observables. We emphasize the role played by antiparticle degrees of freedom in order to control, order by order in the Fock expansion, the scale invariance of physical observables. This nonperturbative framework demands also to use an appropriate regularization scheme. We show why the recently proposed Taylor-Lagrange regularization scheme is a very adequate scheme since it can be implemented very naturally, and from a systematic point of view, in light-front dynamics. As a direct application of this general framework, we settle the basis for a new formulation of chiral effective field theory in the baryonic sector based on light-front dynamics and a Fock decomposition of the state vector.
}

\FullConference{Light Cone 2010: Relativistic Hadronic and Particle Physics\\
		 June 14-18, 2010\\
		 Valencia, Spain}

\begin{document}

The understanding of the structure of relativistic compound systems in nuclear and particle physics is the subject of intense theoretical studies since the discovery of the nucleon structure in the $60$'s. In both domains, the understanding of the properties of bound state systems from the original Lagrangian of quantum chromodynamics (QCD) or from an effective chiral Lagrangian, demands to develop a nonperturbative framework. The interest of light-front dynamics (LFD) to investigate relativistic  bound
states has been advocated for a long time. However, while its use 
in few-body systems has developed rapidly, its application 
to field theory beyond a perturbative framework is only a recent accomplishment.
This is due to the fact that any practical calculation relies on
the truncation of the Fock expansion of the state vector. 

We shall show in this study that we have now all the necessary theoretical tools in order to develop on a large scale a nonperturbative framework to calculate the structure of relativistic compound systems. Our formalism should enable us to control, order by order in the Fock expansion, the scale invariance of physical observables, and hence to make reliable predictions for these observables. 

\section{Covariant formulation of light-front dynamics} \label{cov}
LFD is one of the three forms of dynamics proposed by Dirac in 1949 \cite{dirac} to describe physical systems, bound states as well as scattering states. In this form of dynamics, the state vector describing the system is defined on the surface in four-dimensional space-time given by $t^+=t+z/c$. 
According to Dirac's classification, the ten generators of the
Poincar\'e group, given by space-time translations (four
generators), space rotations (three generators), and Lorentz
boosts (three generators), can be separated into kinematical and
dynamical operators. The kinematical operators leave the
light-front plane invariant and are independent of dynamics, i.e.
of the interaction Hamiltonian of the system, while the dynamical
ones change the light-front position and depend therefore on the
interaction. Among the kinematical operators, one finds, in LFD,
the boost along the $z$ axis. This property is of particular
interest when one calculates electromagnetic observables at high
momentum transfer.
However, the spatial
rotations in the $xz$ and $yz$ planes become dynamical, in
contrast to the case of equal-time dynamics. This is a direct consequence of the violation of rotational invariance caused by the non-invariant definition of the light-front plane orientation.
This violation should be kept under control.

While rotational invariance should be recovered automatically in
any exact calculation, this is not {\em a priori} the case if the
Fock expansion is truncated. In order to control the violation of
rotational symmetry, we shall use the covariant formulation of LFD
(CLFD)~\cite{karm76,cdkm}, which provides a simple, practical, and
very powerful tool in order to describe physical systems. In this formulation, the
state vector is defined on the plane characterized by  the
invariant equation $\omega \cd x=\sigma$, where $\omega$ is an
arbitrary light-like ($\omega^2=0$) four-vector, and $\sigma$ is the light-front time. The  standard LFD
is recovered by considering the
particular choice $\omega=(1,0,0,-1)$. The covariance of our
approach relies on the invariance of the light-front plane
equation under any Lorentz transformation of both $\omega$ and
$x$. This implies in particular that $\omega$ cannot be kept the
same in any reference frame,  as it is the case in the standard
formulation of LFD.

There is of course equivalence between the
standard and covariant forms of LFD in any exact calculation.
Calculated physical observables must coincide in both approaches,
though their derivation in CLFD in most cases is much simpler and
more transparent. Indeed, the relation between CLFD and standard
LFD reminds that between the Feynman graph technique and
old-fashioned perturbation theory.
In approximate calculations, CLFD has a definite advantage
in the sense that it enables a direct handle on the contributions
which violate rotational invariance. These ones depend explicitly
on the orientation of the light-front surface (i.e. on $\omega$)
and can thus be separated covariantly from true physical
contributions. 

The transformation properties of the light-front plane are governed by the four-dimensional momentum and angular momentum operators given by
\begin{eqnarray}
\hat P_\mu &=& \hat P_\mu^0+\hat P_\mu^{int}\ \ \ \mbox{with} \ \ \ \hat P_\mu^{int}=\omega_\mu \int H^{int}(x) \delta(\omega \cd x-\sigma)d^4x \label{P} \ ,\\
\hat J_{\mu \nu} &=& \hat J_{\mu \nu}^0+\hat J_{\mu \nu}^{int}\ \ \ \mbox{with} \ \ \hat J_{\mu \nu}^{int}=\int H^{int}(x) (x_\mu \omega_\nu-x_\nu \omega_\mu)\delta(\omega \cd x-\sigma)d^4x \label{J}\ .
\end{eqnarray}
In these equations, $H^{int}$ refers to the interaction Hamiltonian, while $ \hat P_\mu^0$ ($ \hat J_{\mu \nu}^0$) is the standard free four-momentum (four-angular momentum) operator.
According to the general properties of LFD, we decompose the state vector $\phi(p)$  of a physical
system of momentum $p$ in Fock sectors, with
\begin{equation}
\phi(p)=
\sum_{n=1}^{ \infty} \int dD_n \, \phi_n(k_1,\ldots,k_n;p) \,
\delta^4(k_1+\ldots +k_n-p-\omega \tau_n) \left\vert n
\right>,\label{Fock}
\end{equation}
where $\left\vert n \right>$ is the state containing $n$ free
particles with four-momenta $k_1,\ldots,k_n$ and $\phi_n$'s
are relativistic $n$-body wave functions, the so-called Fock
components. The phase space volume is denoted by $dD_n$.  Note the particular
overall momentum conservation law given by the $\delta$-function.
It follows from the general transformation properties of the
light-front plane $\omega \cd x=\sigma$ under the four-dimensional
translations governed by $\hat P_\mu$ in (\ref{P}). The quantity $\tau_n$ is a measure of how far the
$n$-body system is off the energy shell. It is completely determined by this conservation law and
the on-mass-shell condition for each individual particle momentum.
It is convenient to introduce, instead of the wave functions
$\phi_n$, the vertex functions $\Gamma_n$ defined by
$\Gamma_n =(s_n-M^2)\phi_n$, where $M$ is the physical mass of the bound state and $s_n=(k_1+k_2+\ldots)^2$.
The system of coupled equations for the Fock components of the
state vector can be obtained from the general eigenstate equation \cite{kms_08}
\begin{equation} \label{P2M2}
\hat P^2 \phi(p) = M^2 \phi(p).
\end{equation}
The state vector is finally normalized to $1$.

As follows from the transformation properties of the state vector under the four-dimensional angular momentum (\ref{J}) (the so-called angular condition \cite{cdkm}), the spin structure of the
wave functions $\phi_{n}$ is very simple, since its construction
does not require the knowledge of dynamics. It should incorporate
however $\omega$-dependent components. It is convenient to
decompose each wave function $\phi_{n}$ into invariant amplitudes
constructed from all available  particle four-momenta (including
the four-vector $\omega$!) and spin structures (matrices,
bispinors, etc.). In the Yukawa model for instance (one constituent fermion coupled to scalar bosons), we have for
the one- and two-body components~\cite{kms_08}
\begin{equation}
\label{oneone}
\Gamma_{1}= a_1 \bar{u}(k_1)u(p),\ \ \ 
\Gamma_{2}=\bar{u}(k_1) \left[ b_1  +
b_2\ \frac{M \omega\!\!\!\!\!/ }{\omega \cd p}\right]
u(p),
\end{equation}
since no other independent spin structures can be constructed.
Here $a_1$, $b_1$, and $b_2$ are scalar functions
determined by dynamics, and $k_1$ is the momentum of the constituent fermion. 

\section{The Fock sector dependent renormalization scheme} \label{fock}
\begin{figure}[b,yt,h]
\begin{center}
\includegraphics[width=20pc]{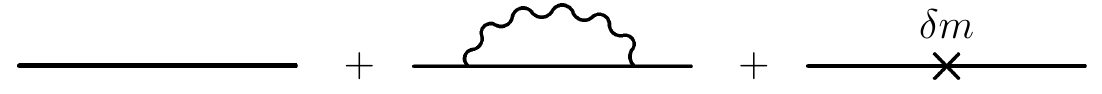}
\caption{Renormalization of the fermion propagator in the second
order of perturbation theory. The fermion mass counterterm is
denoted by $\delta m$. \label{self}}
\end{center}
\end{figure}

The application of LFD to bound state systems in field theory  necessitates to define an appropriate, nonperturbative, renormalization scheme. Indeed,  the full cancellation of
field-theoretical divergences which appear in a given Fock sector
requires to take into account contributions from other sectors. If
even a part of the latter is beyond our approximation, some
divergences may leave uncancelled. This is illustrated in Fig.~\ref{self} for the calculation of
the fermion propagator in second order of perturbation theory:
the self-energy contribution and the fermion mass counterterm
involves two different Fock sectors.
This means that, as a necessary condition for the cancellation of
divergences, any mass counterterm  should be associated  with the
number of particles present (or ``in flight'') in a given Fock
sector. In other words, all mass counterterms must depend on the
Fock sector under consideration, as advocated first in Ref.~\cite{wp}.
This is also true for the renormalization of the bare coupling
constant.

The presence of uncancelled divergences reflects itself in
possible dependence of (approximately) calculated observables on the
regularization parameters, i.e.
approximate physical observables are not anymore scale invariant. This
prevents to make any physical predictions if we cannot control the
renormalization procedure in one way or another. To do that, we have developed
an appropriate renormalization procedure --- the so-called Fock
sector dependent renormalization (FSDR) scheme --- in order to
keep the cancellation of field-theoretical divergences under
permanent control. This scheme amounts to consider a complete set of mass counterterms and bare coupling constants, denoted by  $\delta m^{(n)}$ and  $g_0^{(n)}$ respectively, where $n$ refers to the number of bosons in flight \cite{kms_08}. The mass counterterm is fixed from the solution of the eigenstate equation (\ref{P2M2}), by demanding that the mass of the constituent fermion is identical to the mass of the physical one. The bare coupling constant is fixed from the general, on energy shell, renormalization condition \cite{yukawa}
\begin{equation}\label{recon}
\Gamma_2(s_2=M^2) = g \sqrt{Z_f} \sqrt{Z_b}\ ,
\end{equation}
 where $Z_f$ and $Z_b$ are the field strength renormalization factors of the constituent fermion and boson respectively. When the Fock space is truncated, one should also make sure that all contributions remain within the Fock space.
 
The condition~(\ref{recon}) has two important consequences. The
first one is that the two-body vertex function at $s_2=M^2$ should
be independent of the four-vector $\omega$ which determines the orientation of the
light-front plane. With the spin decomposition~(\ref{oneone}),
this implies that the component $b_2$ at $s_2=M^2$
should be identically zero. While this property is automatically
verified in the case of the two-body Fock space truncation - if
using a regularization scheme which does not violate rotational
invariance - this is not guaranteed for calculations within higher
order truncations.
Indeed, nothing prevents $\Gamma_2$ to be $\omega$-dependent,
since it is an off-shell object, but this dependence must
completely disappear on the energy shell, i.e. for $s_2=M^2$. The truncation of the Fock space may cause some $\omega$-dependence of
$\Gamma_2$ even on the energy shell, which  makes the
general renormalization condition~(\ref{recon}) ambiguous. If
so, one has to insert new counterterms into the light-front interaction
Hamiltonian, which explicitly depend on $\omega$
and cancel the spurious $\omega$-dependence of $\Gamma_2(s_2=M^2)$. Note
that the explicit covariance of CLFD allows to separate the terms
which depend on the light-front plane orientation (i.e. on
$\omega$) from physical contributions, and establish the structure of
these counterterms. This is not possible in ordinary LFD.

The second consequence is that $\Gamma_2(s_2=M^2)$ should be independent of $x$. This is a non-trivial requirement since $\Gamma_2$ does depend in general on two invariant kinematical variables which are usually
chosen as the longitudinal momentum fraction, $x$, of the constituent  fermion, and
the square of its transverse momentum, ${\bf k}_\perp^2$. Hence, if we fix
${\bf k}_{\perp}^2$ from the on energy shell condition $s_2=M^2$, $\Gamma_2$ should be independent of $x$.
Again, this property is verified in the two-body Fock space
truncation, since our equations are
equivalent in this approximation to perturbation theory of order $g^2$. It is not
guaranteed for higher order calculations. In practice, we shall fix
$\Gamma_2(s_2=M^2)$ at some preset value $x^*$ and verify that calculated
physical observables are not sensitive to the choice of $x^*$.

\begin{figure}[ht!]
\begin{center}
\includegraphics[width=17pc]{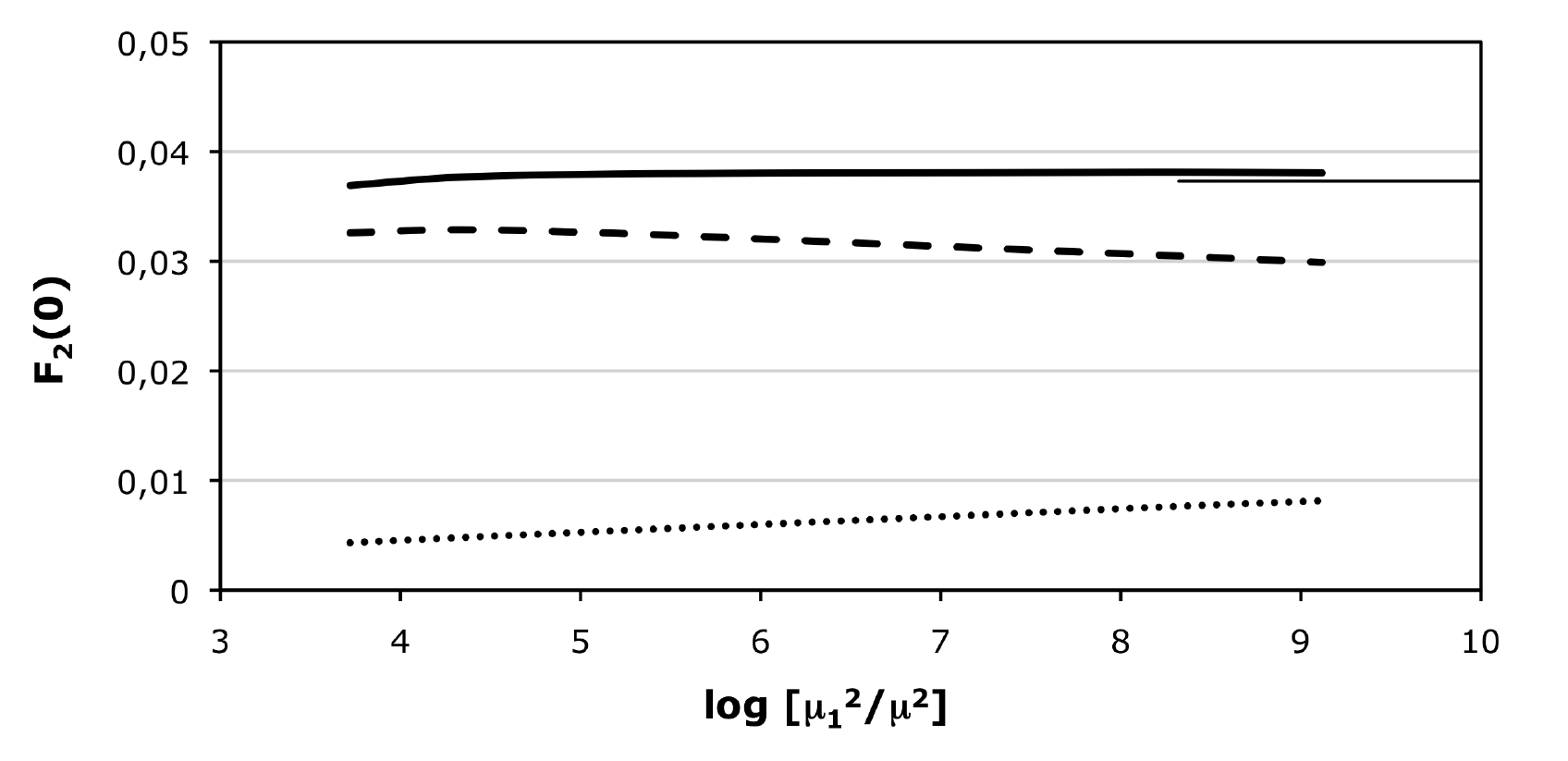}
\includegraphics[width=17pc]{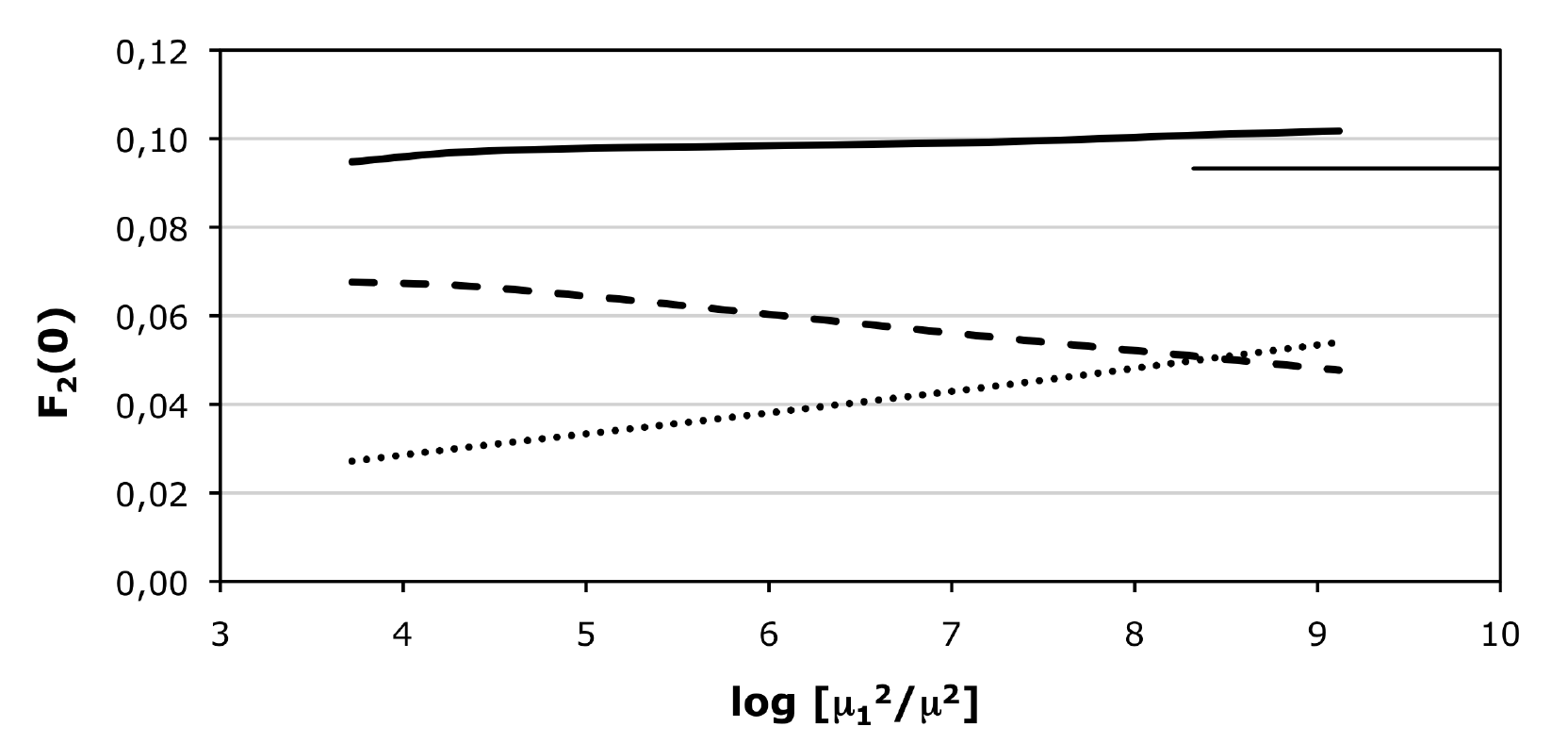}
\end{center}
\caption{The anomalous magnetic moment in the Yukawa model as a
function of the regularization mass  $\mu_1$ (Pauli-Villars boson mass), for two different values of the
coupling constant, $\alpha = 0.2$ (left plot) and $0.5$ (right
plot). The dashed and dotted lines are, respectively, the two- and
three-body contributions, while the solid line is the total
result. The value of the anomalous magnetic moment calculated in
the $N=2$ approximation is shown by the thin line on the right
axis.}\label{amm}
\end{figure}

We show in Figs.~\ref{amm} the anomalous magnetic moment in the Yukawa model~\cite{yukawa}, in the three-body Fock state truncation, for two typical values of the boson-fermion coupling constant. These calculations do not involve antiparticle degrees of freedom. At relatively small values of the coupling constant, they show a rather nice scale invariance, while the calculation at higher values of the coupling constant shows the onset of scale dependence.

\begin{figure}[bth]
\begin{center}
\includegraphics[width=18pc]{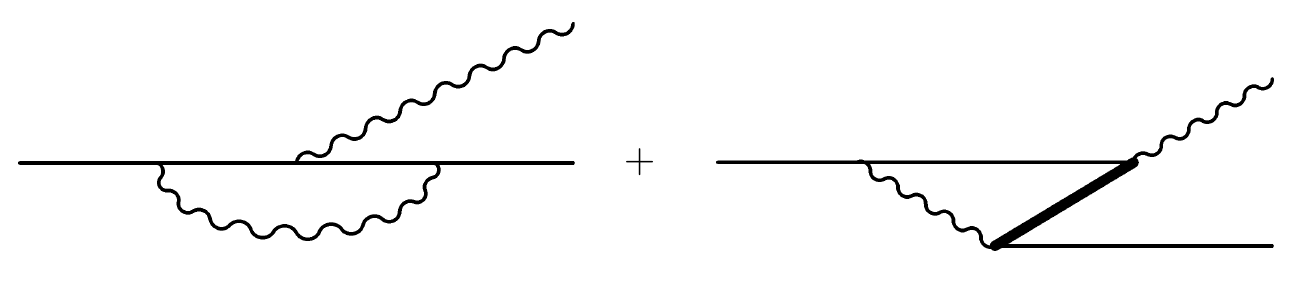}
\end{center}
\caption {Part of the three-body Fock sector contributions to the two-body
vertex. The thick solid line represents an antifermion.}\label{anti}
\end{figure}

We identify this (small) residual scale dependence with the lack of contributions from antiparticle degrees of freedom which should be incorporated explicitly in LFD. To have a better handle on these contributions, it is instructive to calculate
$b_{1,2}(s_2=M^2)$ in perturbation theory. This is done
by calculating the amplitudes of the diagrams shown in
Fig.~\ref{anti}. The contribution of
the left diagram in Fig.~\ref{anti} is
automatically taken into account by the solution of the eigenstate
equation in the $N=3$ Fock state truncation, while the right plot  corresponds to the contribution of fermion-antifermion contributions in leading order. If one calculates the
sum of both contributions~\cite{kms09},
one finds
\begin{eqnarray}
b_{1}^{pert}(s_2=M^2)&=& cte\ , \\
b_{2}^{pert}(s_2=M^2)&=& 0.
\end{eqnarray}
This is a first indication that the expected properties of the
on-shell functions $b_{1,2}$ are indeed recovered when
antifermion degrees of freedom are involved. 
%

\section{The Taylor-Lagrange regularization scheme}
Before choosing an appropriate regularization scheme, it is interesting to come back to the very origin of divergences of loop integrals. These divergences can be traced back to the violation of causality, originating from  ill defined products of distributions  at the same point \cite{aste}. The correct mathematical treatment, known since a long time \cite{EG},  is to consider covariant fields
as operator valued distributions (OPVD), these distributions being applied on test functions with well-defined properties. These considerations lead to the Taylor-Lagrange renormalization scheme (TLRS) \cite{TLRS}. 

If we denote by $f$ the Fourier transform of the test function, a scalar field ${\varphi}(x)$, for instance, will thus write
\begin{equation} \label{conv}
\!\varphi (x)\!=\!\!\int\!\frac{d^{3}{\bf p}}{(2\pi)^{3}}\frac{f(\varepsilon_p^2,{\bf p}^2)}{2\varepsilon_p}
\left[a^\dagger_{\bf p} e^{i{p.x}}+a_{\bf p}e^{-i{p.x}}\right],\ 
\end{equation}
with  $\varepsilon^2_p = {\bf p}^2+m^2$.  
From this decomposition, it is apparent that test functions should be attached to each fermion and boson fields, while
the contraction of two fields (propagator) should be proportional to $f^2$. In LFD, $f$ depends only on ${\bf p}^2$.
These test functions  should have three important properties:

{\it i)} the physical field $\varphi(x)$ in (\ref{conv}) should be independent of the choice of the test function. This may be
achieved if $f$ is chosen 
among the partitions of unity (PU), i.e. if $f$ is build up from a family of functions $\beta_i(X)$ with
\begin{equation}
\sum_{i=1}^L \beta_i(X)=1 \ \ \ \mbox{for} \ \ \ X \in [a,b]\ .
\end{equation}
It is  a function of finite support which is $1$ everywhere except at the boundaries. This 
choice is also necessary in order to satisfy Poincar\'e invariance~\cite{TLRS}. 

{\it ii)} In order to be able to treat in a generic way singular distributions of any type, the test function is chosen as a super regular test function (SRTF), i.e. a function which vanishes, as well as all its derivatives, at  the ultraviolet (UV) and the infrared (IR) boundaries. 

{\it iii)} The boundary conditions of the test function - which is assumed  for simplicity to depend on a  one dimensional variable $X$ - should embody a scale invariance inherent, in the UV domain for instance,  to the limit $X \to \infty$ since in this limit $\eta^2 X$ also goes
 to $\infty$, where $\eta^2$ is an arbitrary dimensionless scale. This can  be done by considering a
running boundary condition for the test function, i.e. a boundary condition which depends on the variable $X$ according to
\begin{equation} \label{running}
f(X \ge H(X)) = 0 \ \ \ \mbox{ for} \ \ \ \ H(X)\equiv \eta^2 X g(X)\ .
\end{equation}

\begin{figure}[bth]
\begin{center}
\includegraphics[width=12pc]{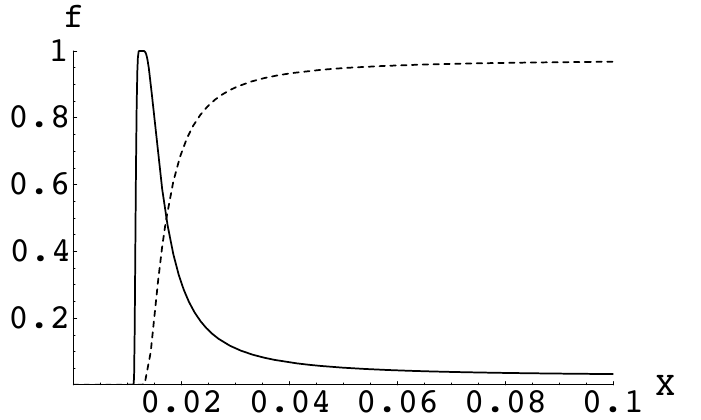}
\includegraphics[width=12pc]{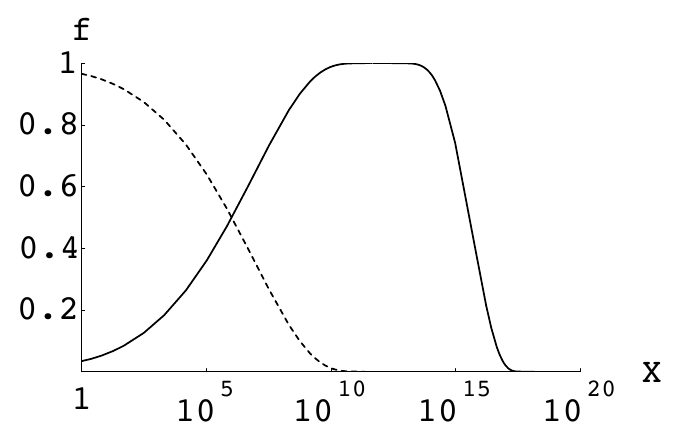}
\end{center}
\caption {Construction of a partition of unity with running support. The left curve shows the IR domain, while the right curve shows the UV domain.}\label{TLRS}
\end{figure}

This condition defines a maximal value, $X_{max}$, with $f(X_{max})=0$.
A typical
example of  the fuction $g(X)$ is given in Ref.~\cite{TLRS}.   This running condition is equivalent to having an ultra-soft cut-off, i.e. an infinitesimal drop-off of the test function in the asymptotic limit, the rate of drop-off being governed by the arbitrary scale $\eta^2$. 
A similar scale invariance is also present in the IR domain. An example of test function constructed from two elementary functions, with a running condition, is shown on Fig.~\ref{TLRS}~\cite{TLRS}

With these properties, the TLRS can be summarized as follows, in the UV domain for instance. Starting from a general amplitude $\cal A$ written for simplicity in a one dimensional space as
\begin{equation} \label{cala}
{\cal A} = \int_0^\infty dX \ T(X) \ f(X) \ ,
\end{equation}
where $T(X)$ is a singular distribution in the UV domain,
we apply the following general Lagrange formula to $f(X)$, after separating out an intrinsic scale $a$ from the (running) dynamical variable $X$
\begin{equation} \label{faX}
f(aX)=-\frac{X}{a^k k!}\int_a^\infty\frac{dt}{t} (a-t)^k \partial_X^{k+1}\!\left[ X^k f(Xt)\right].
\end{equation}
This Lagrange formula is valid for any order $k$, with $k>0$, since $f$ is chosen as a SRTF and it is therefore equal to its Taylor remainder for any $k$. After integration by part in (\ref{cala}), and using (\ref{faX}), we can express the amplitude $\cal A$ as
\begin{equation} \label{afin}
{\cal A} = \int_0^\infty dX \ \widetilde T(X) \ f(X) \ ,
\end{equation}
where $\widetilde T(X)$ is the so-called extension of the singular distribution $T(X)$. In the limit $f \to 1$  obtained by letting $g(X) \to 1^-$, it is given by \cite{TLRS}
\begin{equation} \label{Tex}
\widetilde T(X)\equiv\frac{(-X)^{k}}{a^k k!} \partial_X^{k+1} \left[ X T(X)\right] \int_a^{\eta^2} \frac{dt}{t} (a-t)^k \ .
\end{equation}
The value of $k$ in (\ref{Tex}) corresponds to the order of singularity of the original distribution $T(X)$ \cite{TLRS}.  The extension of $T(X)$ is no longer singular due to the derivatives in (\ref{Tex}),  so that we can safely perform the limit $f \to 1$ in (\ref{afin}).
This amplitude is well defined but depends on the arbitrary dimensionless scale $\eta^2$. This scale is the only remnant of the presence of the test function.
Note that we do not need to know the explicit form of the test function in the derivation of the extended distribution $\widetilde T(X)$. 
We only rely on its mathematical properties  and on the running construction of the boundary conditions. The extension of singular distributions in the IR domain can be done similarly. 

The use of TLRS for the calculation of the state vector of compound systems within CLFD is very natural. Since each vertex function $\Gamma_n$ is attached to one fermion and $n-1$ boson lines, it will be multiplied at least by $f({\bf k}_1^2) \ldots f({\bf k}_n^2)$. We can thus redefine $\Gamma_n$ to include implicitly these test functions:
\begin{equation}
\Gamma_n ({\bf k}_1 \ldots {\bf k}_n) \to \bar \Gamma_n ({\bf k}_1 \ldots {\bf k}_n)=  \Gamma_n({\bf k}_1 \ldots {\bf k}_n) f({\bf k}_1^2) \ldots f({\bf k}_n^2) \ .
\end{equation}
Since the $f's$ are SRTF, this implies that any $\bar \Gamma_n$ is also a super regular function with respect to all momenta. We can thus apply the Lagrange formula to all loop calculations, and derive the extension of all singular distributions along the lines detailed in the previous section \cite{TLRS}.
 
 It is instructive to calculate the extension of the singular distribution $T(X)=1/(X+a)$. Using (\ref{faX}) and (\ref{afin}) with $k=0$, one finally arrives to the following extension \cite{TLRS2}
 \begin{equation}
 \widetilde T(X) = \frac{1}{X+a}- \frac{1}{X+\eta^2}\ .
 \end{equation}
 This form is reminiscent of a Pauli-Villars subtraction, with however two important differences. This subtraction involves the arbitrary scale $\eta^2$. We thus do not have to perform any infinite mass limit. Moreover, this extension results from the application of TLRS to physical contributions, and we do not have to introduce any ad-hoc Pauli-Villars particles.

\section{Light front chiral effective field theory}
It is now well recognized that the properties of QCD at low energies can be understood in terms of an effective Lagrangian based on the (approximate) chiral symmetry of strong interactions, and its spontaneous breakdown. The physical properties of pions within chiral perturbation theory (CPT) are now well reproduced in a meaningfull power expansion. This originates from the fact that the pion has zero mass in the chiral limit. This is however not the case in the nucleon sector, and all momentum scales are {\it a-priori} involved in the calculation of nucleon properties.

\begin{figure}[btph]
\begin{center}
\includegraphics[width=18pc]{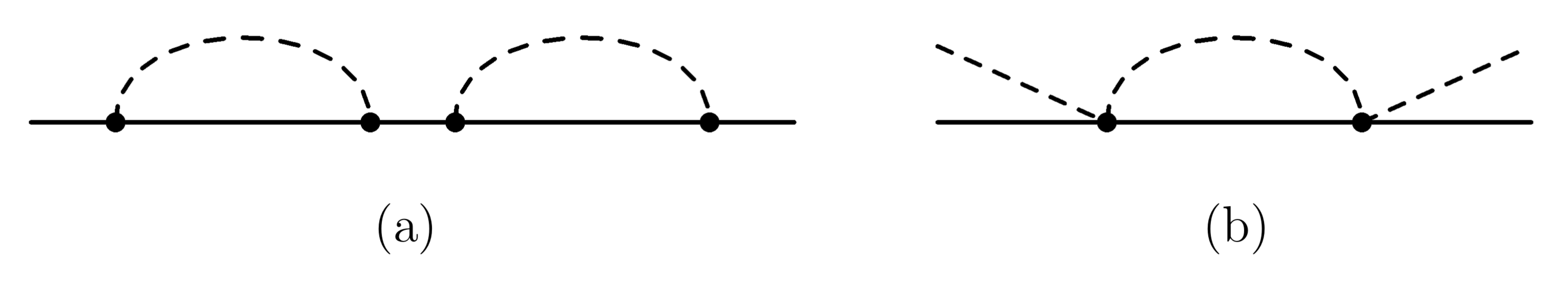}
\caption{Iteration of the self-energy contribution in the first
order of perturbation theory (a); irreducible contribution to the
bound state equation (b). Dashed lines represent
pions.\label{self_gen}}
\end{center}
\end{figure}

Since  the pion mass is zero in the chiral limit, any calculation
of $\pi N$ systems demands a relativistic framework to get, for
instance, the  right analytical properties of the physical
amplitudes. The calculation of compound systems, like a physical
nucleon composed of a bare nucleon coupled to many pions, relies
also on a nonperturbative eigenstate equation. While the mass of
the system can be determined in leading order from the iteration
of the $\pi N$ self-energy calculated in the first order of
perturbation theory, as indicated in Fig.~\ref{self_gen}(a), this
is in general not possible, in particular, for $\pi N$ irreducible
contributions, as shown in Fig.~\ref{self_gen}(b).

While there is not much freedom, thanks to chiral symmetry, for
the construction of the effective Lagrangian in CPT in terms of the pion field ${\bf \pi}$ --- or more
precisely in terms of the U field defined by $U=e^{i {\bf
\tau}.{\bf \pi}/f_\pi}$, where $f_\pi$ is the pion decay constant
and $ \bf \tau$ are the Pauli matrices, --- one should settle an
appropriate approximation scheme in order to calculate nucleon
properties. Up to now, two main strategies have been adopted. The
first one is to force the bare (and hence the physical) nucleon
mass to be infinite, in heavy baryon chiral perturbation
theory~\cite{manohar}. In this case, by construction, an expansion
in characteristic momenta can be developed. The second one is to
use a specific regularization scheme~\cite{IR} in order to
separate contributions which exhibit a meaningful power expansion,
and hide the other parts in appropriate counterterms. In both
cases however, the explicit calculation of baryon properties
relies on an extra approximation in the sense that physical
amplitudes are further calculated by expanding the effective
Lagrangian, denoted by ${\cal L}_{CPT}$, in a finite number of
pion fields in perturbation theory.

The general framework we have developed above is particularly
suited to deal with these requirements in the nucleon sector. This leads to the
formulation of light-front chiral effective field theory
(LF$\chi$EFT)~\cite{LF_jf} with a specific effective Lagrangian
${\cal L}_{eff}$. The decomposition of the state vector in a
finite number of Fock components implies to consider an effective
Lagrangian which enables all possible elementary couplings between
the pion and nucleon fields compatible with the Fock space
truncation. This is indeed easy to achieve in CPT since each derivative of the $U$ field involves one
derivative of the pion field. In the chiral limit, the chiral
effective Lagrangian of order $p$, ${\cal L}_{CPT}^p$, involves
$p$ derivatives and therefore at least $p$ degrees of the pion
field. In order to calculate the state vector in the $N$-body
approximation, with one fermion and $(N-1)$ pions, one has
therefore to include contributions up to $2(N-1)$ pion fields in
the effective Lagrangianv (for pions in the initial and final states). We thus
should calculate the state vector in the $N$-body Fock state truncation
with an effective Lagrangian denoted by ${\cal L}_{eff}^N$ with
\begin{equation}
{\cal L}_{eff}^N={\cal L}_{CPT}^{p=2(N-1)}\ .
\end{equation}

While the effective Lagrangian in LF$\chi$EFT can be mapped out to
the CPT Lagrangian of order $p$, the calculation of the state
vector does not rely on any momentum decomposition. It relies only
on an expansion in the number of pions in flight at a given
light-front time. In other words, it relies on an expansion  in
the fluctuation time, $\tau_f$, of such contributions. From
general arguments, the more particles we have at a given
light-front time, the smaller the fluctuation time is. At low
energies, when all processes have characteristic interaction times
larger than $\tau_f$, this expansion should be meaningful.

Moreover, the use of TLRS enables to perform systematic nonperturbative calculations in the general framework developped in Secs.~\ref{cov} and \ref{fock}, contrarily to the regularization schemes used up to now. This finite regularization scheme is also of particular interest since it does not involve any large momentum or energy scale which may break chiral symetry.

It is interesting to illustrate the general features of
LF$\chi$EFT calculations. At order $N=2$, we already have to deal
with irreducible contributions, as shown in
Fig.~\ref{self_gen}(b). It leads to a non-trivial renormalization of
the coupling constant. The calculation at order $N=3$ incorporates
explicitly contributions coming from $\pi \pi$ interactions ($\sigma$ and $\rho$ resonances), as
well as low energy $\pi N$ resonances, like the $\Delta$ or
Roper resonances.
 We
can generate therefore all these resonances in the intermediate
state without the need to include them explicitly, provided the
effective Lagrangian has the right dynamics to generate these
resonances.

\section{Perspectives}
The general framework we presented above to study the structure of relativistic compound systems in light-front dynamics in a nonperturbative way relies on three main advances:
{\it i)} the construction of a covariant formulation of light-front dynamics \cite{karm76, cdkm} in order to control any violation of rotational invariance;
{\it ii)}  the development of an appropriate renormalization scheme --- the so-called Fock sector dependent renormalization scheme --- to deal with the truncation of the Fock expansion \cite{kms_08};
{\it iii)}  the use of an appropriate regularization scheme --- the so-called Taylor-Lagrange regularization scheme --- very well adapted to systematic calculations in light-front dynamics \cite{TLRS2}.

These advances should enable us to have a predictive framework order by order in the Fock expansion. We shall complete in the future this description by considering physical systems involving spontaneous symmetry breaking. It is known that these systems can be described in light-front dynamics by the consideration of zero-mode contributions, in the $\lambda \phi^4$ theory in $1+1$ dimension for instance \cite{hksw}. Their full calculation in $3+1$ dimensions within the general framework presented above remains to be done.


\end{document}